# The rate of decline of CD4 T-cells in people infected with HIV

Brian G. Williams,[1] Eline L. Korenromp,[2] Eleanor Gouws,[3] and Christopher Dye[4]

1. South African Centre for Epidemiological Modelling and Analysis (SACEMA), Stellenbosch, South Africa
2. The Global Fund to Fight AIDS, Tuberculosis and Malaria, and Department of Public Health, Erasmus MC, University Medical Centre Rotterdam, Rotterdam, The Netherlands
3. Policy, Evidence and Partnerships Department, Joint United Nations Programme on HIV/AIDS, Geneva, Switzerland
4. World Health Organization, Geneva, Switzerland

Correspondence to: Brian Williams at williamsbg@me.com.

## Abstract

In people infected with HIV the RNA viral load is a good predictor of the rate of loss of CD4 cells at a population level but there is still great variability in the rate of decline of CD4 cells among individuals. Here we show that the pre-infection distribution of CD4 cell counts and the distribution of survival times together account for 87% of the variability in the observed rate of decline of CD4 cells among individuals. The challenge is to understand the variation in CD4 levels, among populations and individuals, and to establish the determinants of survival of which viral load may be the most important.

## Introduction

A recent study by Rodríguez et al. [1], using data from three cohorts of HIV-positive people in the United States of America, showed that RNA viral load in patients presenting with HIV infection is positively correlated with the rate of loss of CD4 T-cells at a population level ($n = 1289$, $p < 0.0001$; our estimate), confirming the earlier results of Mellors et al. [2]. However, the RNA viral load only explained 4% (2% to 6%, 95% confidence limits,) of the variability in the rate at which CD4 T-cells declined among individual people. Rodríguez et al. [1] asked what other factors are likely to drive the decline in CD4 T-cells and account for the remaining 96% of the variability and noted the implications of these findings for treatment decisions and for understanding the pathogenesis of HIV. In subsequent letters Gottlieb et al. [3] and Lima et al. [4] responded that the paper by Rodríguez et al. [1] challenges the utility of using early viral load measurements to predict disease outcome in individuals and the rate at which CD4 cells are lost and suggested the use of more sophisticated techniques for the analysis of the data.

We have recently published an epidemiological model of CD4 cell count decline among populations of HIV-infected adults in Zambia and South Africa [5]. Our model takes as input, for a particular population, the frequency distribution of CD4 T-cell counts among HIV-negative adults and the distribution of survival times after HIV-infection. We assume that there is a 25% decline in CD4 cell counts in the few weeks immediately following infection [6-8] and that the distribution of survival times after HIV-infection is independent of the initial CD4 cell count. Using this model we were able to predict the distribution of CD4 counts in the corresponding HIV-positive population with no free parameters apart from the normalization to match the size of the sample. Applied to data from South African and Zambian populations, the model explained 82% and 76%, respectively, of the variability in the distribution of CD4 cell counts among HIV-infected adults; the remaining variability was entirely accounted for by the stochastic sampling errors in the data. Here we use our model to predict the distribution of the rate of decline of CD4 cell counts in the population studied by Rodríguez et al. [1].

## Methods

Rodríguez et al. [1] used data from people infected with HIV enrolled in the Research in Access to Care for the Homeless Cohort (REACH), the San Francisco Men's Health Study (SFMHS) cohort and the Multicenter AIDS Cohort Study (MACS) cohort. Because they did not present data on the distribution of CD4 cell counts among the corresponding HIV-negative populations we used the distribution among patients in their study in the lowest category of presenting RNA (less than 0.5/μL, median 0.084/μL), who also have the highest CD4 cell counts (median 654/μL, inter-quartile range 474−864/μL). We assume that the distribution of CD4 cell counts in this group of patients represents the distribution in HIV-infected adults immediately after the initial decline in the few weeks following infection. A log-normal curve gave a good fit to the distribution of CD4 cell counts for HIV-negative adults in the South African and Zambian populations [5] and we fitted a log-normal curve to the median and inter-quartile range of CD4 counts given by Rodríguez et al. [1]. Based on data from the CASCADE study [9], we use a Weibull function with a mean of ten years and a shape parameter of 2.25 [5] for the distribution of survival after HIV infection.

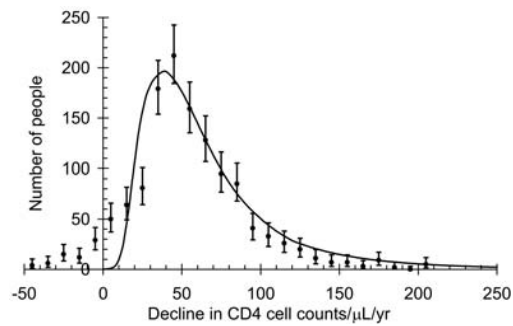

Figure 1. The distribution of the rate of decline of CD4 T-cells in HIV-infected adults as reported by Rodríguez et al. [1]. Dots and error bars are the data summed over the whole sample; the line is the distribution predicted using the model described in the text [5]. The vertical axis gives the number of people with HIV in each category of CD4 decline (/10μL/year).

## Results

The distribution of the rate at which CD4 cell counts decline as predicted by our model is compared to the data published by Rodríguez and et al. [1], summed over all the participants in their study, in Figure 1. There is good agreement between the predicted and observed distributions of the rate at which CD4 cell counts decline (Figure 1) apart from a small number of people whose viral load was either falling slowly or increasing (points close to zero or the left of the vertical axis in Figure 1.) The model accounts for 87% of the variability in the data.

## Conclusions

The predicted distribution depends on the variability in the initial, pre-infection CD4 cell counts and the variability in the survival after infection with HIV. The question to be asked is: what determines each of these? CD4 cell counts vary widely among HIV-negative adults in association with a range of genetic, biological and physiological effects including smoking, physical exercise, and chronic infections [10]. In individual people significant fluctuations may be recorded between readings taken only a few weeks apart, due to diurnal rhythms, acute infections, stress, exercise or the reproducibility of test results [10]. Any additional random variation, over and above that explained by the model, would further broaden the measured distribution and could account for the slow rates of decline, and even increases, in CD4 cell counts observed in a few of the patients as shown in Figure 1. The distribution of survival after HIV infection and without ART is well characterized and varies most obviously with the age at infection [9] and with viral load [2] although the detailed mechanisms are not well understood.

Rodríguez et al. [1] challenge the scientific community to explain the substantial variation in the observed rates at which CD4 cell counts decline in people infected with HIV. While RNA viral load only explains 4% of the variability in the rates at which individual CD4 cell counts decline, our model shows that most of the variability in the rate at which CD4 cell counts decline can be explained by the distribution of the pre-infection CD4 cell counts and the distribution of survival after infection. The reality must be considerably more complex than suggested by our simple model where we take the distribution of survival times as given leaving the relationship between viral load and survival an open question. If it were the case that individual RNA viral load completely determines the individual survival, so that the distribution of the former completely determines the distribution of the latter, the variation arising from the variability in initial CD4 cell counts would remain and the model would be consistent with the claim by Mellors et al. [2] While our analysis does not fully answer the questions raised in the paper by Rodríguez et al. [1] or in the ensuing correspondence [3, 4] we believe that it helps to focus the questions.

## Acknowledgements

Financial support was received from: Erasmus MC, University Medical Centre Rotterdam (Van Rijn fellowship to E.L.K.). This article is the work of the authors and is not a policy statement of the World Health Organization, the Joint United Nations Programme on HIV/AIDS, or The Global Fund to Fight AIDS, TB and Malaria.